\documentclass[12pt, a4paper]{article}
\usepackage{xcolor}
\usepackage[pdftex,colorlinks=true]{hyperref}
\definecolor{darkblue}{rgb}{0,0,.6}
\hypersetup{citecolor=darkblue, linkcolor=darkblue, urlcolor=darkblue}
\usepackage{amsmath, caption, orcidlink}
\usepackage[top=1.4in, bottom=1.4in, left=1.2in, right=1.2in]{geometry}
\usepackage{nicematrix}
\usepackage{makecell} 

\DeclareCaptionStyle{italic}[justification=centering]{labelfont={bf}, textfont={it}, labelsep=colon}
\usepackage{graphicx,psfrag,epsf}
\usepackage{enumerate}
\usepackage{natbib}
\usepackage{url} 
\usepackage{booktabs, subfig, bm, paralist, mathpazo, tikz, todonotes, longtable, microtype, dsfont, rotating} 
\newcommand{\blind}{0}

\graphicspath{{plots/}}
\DeclareMathOperator*{\argmin}{\arg\!\min}
\newsavebox\CBox

\definecolor{a0}{rgb}{0.0, 0.5, 0.0}
\definecolor{bistre}{rgb}{0.24, 0.17, 0.12}
\definecolor{amethyst}{rgb}{0.6, 0.4, 0.8}
\definecolor{blue-violet}{rgb}{0.54, 0.17, 0.89}
\definecolor{Rcolor}{RGB}{150,160,190}
\definecolor{blush}{rgb}{0.87, 0.36, 0.51}
\definecolor{brightturquoise}{rgb}{0.03, 0.91, 0.87}
\definecolor{burntorange}{rgb}{0.8, 0.33, 0.0}
\date{}
\AtBeginDocument{}
\newcommand{\X}{\mathcal{X}}
\newcommand{\Y}{\mathcal{Y}}

\newcommand{\Rlogo}{\protect\includegraphics[height=1.8ex,keepaspectratio]{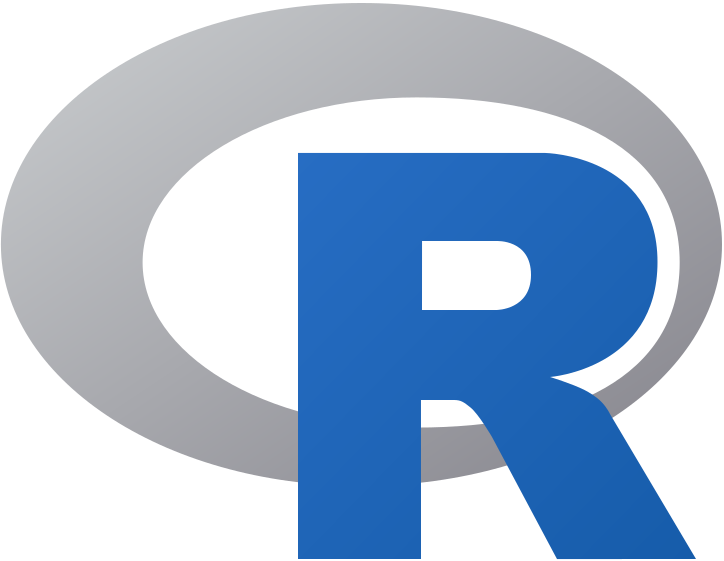}}

\begin{document}

\def\spacingset#1{\renewcommand{\baselinestretch}
{#1}\small\normalsize} \spacingset{1}

\if0\blind
{
  \title{\bf Change-point detection in functional time series: \hbox{Applications to age-specific mortality and fertility}}
  \author{Han Lin Shang\thanks{Postal address: Department of Actuarial Studies and Business Analytics, Level 7, 4 Eastern Road, Macquarie University, Sydney, NSW 2109, Australia; Telephone: +61(2) 9850 4689; Email: hanlin.shang@mq.edu.au.} \orcidlink{0000-0003-1769-6430}
  \hspace{.2cm}\\
    Department of Actuarial Studies and Business Analytics \\
    Macquarie University
    }
  \maketitle
} \fi

\if1\blind
{
  \title{\bf Change-point detection in functional time series: \hbox{Applications to age-specific mortality and fertility}}
  \maketitle
} \fi

\bigskip

\begin{abstract}

We consider determining change points in a time series of age-specific mortality and fertility curves observed over time. We propose two detection methods for identifying these change points. The first method uses a functional cumulative sum statistic to pinpoint the change point. The second method computes a univariate time series of integrated squared forecast errors after fitting a functional time-series model before applying a change-point detection method to the errors to determine the change point. Using Australian age-specific fertility and mortality data, we apply these methods to locate the change points and identify the optimal training period to achieve improved forecast accuracy.
\\

\noindent Keywords: dynamic functional principal component analysis; integrated squared forecast error; long-run covariance function; structural breaks.

\end{abstract}

\spacingset{1.5}

\section{Introduction}\label{sec:1}

In many developed countries, increases in longevity and an aging population have raised concerns about the sustainability of pensions, healthcare, and aged-care systems \citep[e.g.,][]{Coulmas07, OECD13}. These concerns have sparked significant interest among government policymakers, planners, demographers, and actuaries in accurately modeling and forecasting age-specific demographic rates.

Many statistical methods have been proposed for forecasting age-specific mortality rates \citep[see reviews by][]{CDE04, Booth06, BT08, GK08, SBH11, TB14}. Among these, a significant milestone in demographic forecasting is the work of \cite{LC92}. They implemented a principal component method to model age-specific mortality rates, extracting a single time-varying index of the level of mortality rates. Forecasts are then obtained from this index using a random walk with drift.

The Lee-Carter (LC) model is a factor model with age and period effects. In the demographic literature, several extensions have been proposed. From a time-series matrix perspective, \cite{BMS02}, \cite{RH03}, \cite{CBD06}, and \cite{CBD+09} suggest using more than one component in the LC method to model mortality. \cite{RH06} introduce the age-period-cohort LC method, while \cite{Plat09} extend the LC model by incorporating age dependencies. Bayesian techniques for LC model estimation and forecasting have been considered by \cite{GK08} and \cite{WSB+15}. \cite{HH09} follow a generalized linear model, resulting in models with a structure similar to the LC model but with a generalized error structure. From a time series of functions perspective, \cite{HU07} propose a functional data model utilizing nonparametric smoothing and higher-order principal components. In contrast, \cite{LL05}, \cite{HBY13}, \cite{SSB+16}, and \cite{Shang16} model age-specific mortality for multiple populations jointly. For a comprehensive review of the LC model, refer to the recent survey article by \cite{BCB22}.


Modeling mortality consists of two parts:
\begin{inparaenum}
\item[(1)] estimate a mortality model using historical data;
\item[(2)] forecasting the time-dependent parameters from the estimated model.
\end{inparaenum}
The current literature on mortality modeling has primarily focused on creating more extensive mortality models, largely addressing the first part of the process, while the modeling of time-dependent effects in mortality models has not been extensively explored in recent literature, with the exception of \cite{BMS02}. Typically, time-dependent effects (i.e., period and cohort effects) are modeled using autoregressive integrated moving average (ARIMA) models. However, these time-dependent effects may exhibit one or multiple change points. \cite{BMS02} consider a structural break method applied to the first set of estimated principal component scores. This regression-based method regresses the set of scores against the year. Although the first set of scores explains the most variation in the data, it does not adequately justify the reflection in the scores of a change point in the original data. In the implementation of \citeauthor{BMS02}'s \citeyearpar{BMS02} method, a minimum number of years is set in the fitting period, which is arbitrarily chosen to be 20 years by default. This choice limits the detection of change points.

Various exogenous factors influence mortality over time. Severe short-term events, such as pandemics, wars, and natural disasters, contribute significantly to mortality risks, often resulting in more deaths than expected and introducing spikes in age-specific mortality with lasting effects. In \cite{BMS02}, the observed spike is not due to a crisis but rather a sustained effort to reduce mortality from cardiovascular disease, which included new treatments and lifestyle changes \citep{MV02}. Similarly, government policy interventions aimed at encouraging childbearing to address low birth rates can have an impact on age-specific fertility rates in many developed countries. Identifying these spikes, known as change points, is crucial in modeling mortality and fertility.

Time series are sequences of measurements over time that describe the behavior of systems. These behaviors can change due to external events and/or systematic internal changes in dynamics distribution. When structural changes are present, standard ARIMA models may not adequately capture time-dependent effects, leading to inconsistent results during the calibration period. \cite{MLC11} propose a regime-switching model with two regimes for different segments of time-varying principal component scores, calibrating the LC model on US data. These two regimes can have different means and variances. \cite{Hainaut12} extends the regime-switching model to \citeauthor{RH03}'s \citeyearpar{RH03} model and finds a significant increase in log-likelihood.

While age is often observed discretely, treating age as a continuous variable may offer advantages. Using interpolation or smoothing, we can obtain $\X_t(u)$ for $t=1,\dots,n$ where $u$ denotes a continuous variable, such as age. \cite{HS09} and \cite{Shang16} apply functional data analysis to model and forecast age-specific demographic rates for a given year as a continuous function, which can later be discretized into discrete ages at any sampling interval. \cite{HU07} emphasize that the main advantage of functional data analysis is the incorporation of smoothing techniques into the modeling and forecasting of functions, allowing for the natural handling of missing or noisy data. Another advantage is the ability to consider derivatives, a by-product of treating the data as functions \citep[see, e.g.,][]{Shang18, HS22}.

The conditions under which functions are observed over time may undergo changes. There has been increasing interest in detecting and estimating change points in the mean function over time and subsequently estimating locations of change points if they exist. One school of thought is to first reduce the infinite dimension of functional data to a finite dimension via the classic functional principal component analysis, then use the detection method developed for multivariate time series to identify breaks \citep[e.g.,][]{AGH+09, BGH+09, ZSH+11, AK12}. The other is a fully functional detection method without preliminary dimension reduction \citep[e.g.,][]{HKR14, STW16, ARS18}, which avoids possible information loss caused by the dimension reduction.

The primary contribution of this paper is to introduce two change point detection methods designed for functional time series and to apply them to age-specific demographic rates. The first method, developed by \cite{ARS18}, utilizes a (scaled) functional cumulative sum statistic to pinpoint the change point. The second method, developed by \cite{SCS22}, applies a standard structural break method to a time series of integrated squared errors obtained after fitting a functional time-series model. The first method aims to identify the largest gap in the cumulative sum where a change point may occur. The second method transforms the change point detection into a forecasting exercise where a detection method can be applied to the univariate time series of residuals.


The remainder of this paper is organized as follows. Section~\ref{sec:2} presents the motivating dataset, which comprises Australian age-specific demographic rates. In Section~\ref{sec:3}, we introduce two change-point detection methods. Via a series of simulation studies, the comparison of finite-sample performance between the two methods is presented in Section~\ref{sec:6}. These methods are then applied to identify change points in the Australian age-specific mortality and fertility rates in Section~\ref{sec:4}. Section~\ref{sec:5} concludes with a summary of findings and discusses potential avenues for extending the methodology further.

\section{Data sets}\label{sec:2}

\subsection{Australian age-specific mortality rates}\label{sec:2.1}

\begin{figure}[!htb]
\centering
\includegraphics[width=8.4cm]{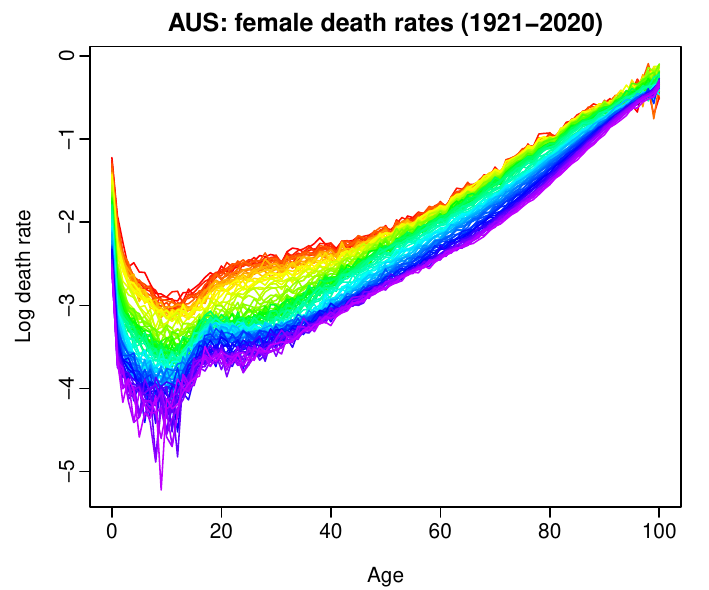}
\quad
\includegraphics[width=8.4cm]{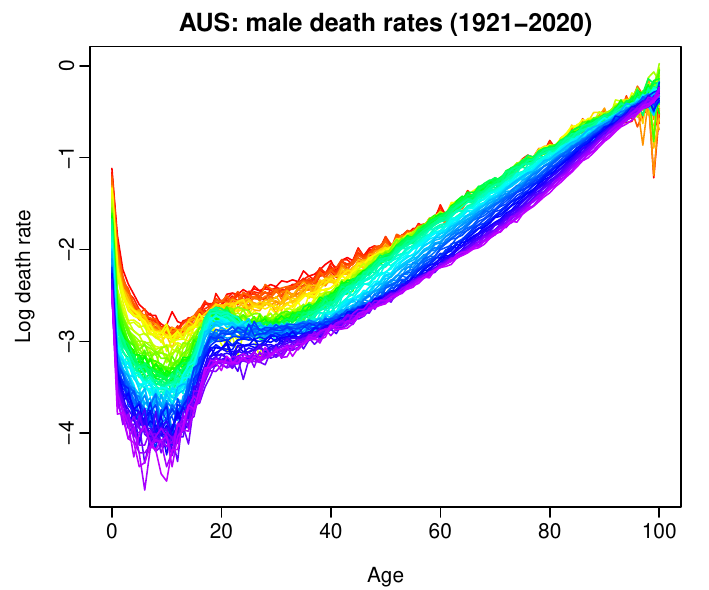}
\caption{\small{Rainbow plots illustrating Australian age- and sex-specific mortality rates from ages 0 to 100+ between 1921 and 2020.}}\label{fig:1}
\end{figure}

We analyze Australian age- and sex-specific mortality rates spanning from 1921 to 2020, obtained from the \cite{HMD24}. These rates represent the ratio of death counts to population exposure in each respective year and age group (based on one-year intervals). Our study covers age groups from 0 to 99 in single years, with the final group encompassing ages 100 and above. In demographic studies, age-specific mortality rates are often modeled and forecasted using logarithmic transformations for ease of interpretability. To illustrate this, we present rainbow plots for $\log_{10}$ mortality rates \citep[see also][]{YSC22} in Figure~\ref{fig:1}, where earlier data are depicted in red and more recent data in purple.

Applying the functional KPSS test introduced by \cite{HKR14}, we obtain $p$-values of 0.007 and 0.009 for the female and male $\log_{10}$ mortality rates, respectively. Based on these $p$-values, we conclude that both series exhibit non-stationarity in their mean.

\subsection{Australian age-specific fertility rates}\label{sec:2.2}

\begin{figure}[!htb]
\centering
\includegraphics[width=11.5cm]{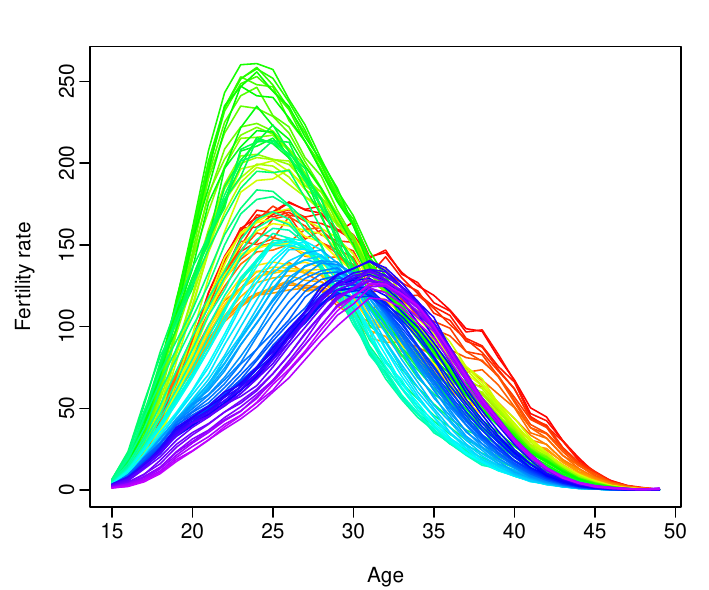}
\caption{\small{Rainbow plot illustrating Australian age-specific fertility rates between ages 15 and 49 from 1921 to 2021.}}\label{fig:2}
\end{figure}

Similar to other developed countries, fertility rates in Australia have declined significantly, dropping from 66 per 1,000 in 2007 to 56 per 1,000 in 2020. We examine annual Australian fertility rates spanning from 1921 to 2021 for ages 15--49, sourced from the Australian Bureau of Statistics (\url{https://www.abs.gov.au/statistics/people/population/births-australia/latest-release}). These rates are defined as the number of live births during each calendar year per 1,000 female residents of the same age on 30 June. Figure~\ref{fig:2} displays the time series of these age-specific fertility rates.

Applying the functional KPSS test, we obtain a $p$-value of 0.045 for the age-specific fertility rates. Based on this result, we conclude that the series is non-stationary in the mean.

\section{Change-point detection methods}\label{sec:3}

In the functional time-series literature, there are three types of detection techniques.
\begin{enumerate}
\item[(1)] Dimension reduction approach: This method involves reducing the infinite dimensionality of functional data to a finite dimension using functional principal component analysis. Subsequently, detection methods designed for multivariate data are applied to identify breaks \citep[see, e.g.,][]{AGH+09}.
\item[(2)] Fully functional method: To preserve information without dimension reduction, fully functional methods are employed \citep{ANH15}.
\item[(3)] Regression-based approach: Another approach involves regression-based methods, such as those discussed by \cite{SCS22}, which are applied to a univariate time series of forecasting errors.
\end{enumerate}

\subsection{Fully functional detection method}\label{sec:3.1} 

Denote $[\X_1(u), \X_2(u), \dots,\X_n(u)]$ as a time series of demographic curves, such as age-specific mortality or fertility rates. These observations are generated from a model
\begin{equation*}
\X_t(u) = \mu(u) + \delta \mathds{1}\{t>\eta^{*}\}+\epsilon_t(u),\quad t=1,\dots,n,
\end{equation*}
where $\eta^{*}=\lfloor \vartheta n\rfloor$, with $\vartheta\in (0,1)$ representing the unknown breakpoint location, $\mathds{1}\{\cdot\}$ denotes a binary indicator function, and $\mu(u)$ and $\epsilon_t(u)$ denote the mean and error terms, respectively.

\cite{ANH15} propose a hypothesis test to examine whether $\delta = 0$. The test statistic is constructed based on a scaled functional cumulative sum statistic:
\begin{equation}
S_{n, \eta}^{0}(u)= \frac{1}{\sqrt{n}}\left(\sum^{\eta}_{t=1}\X_t(u) - \frac{\eta}{n}\sum^n_{t=1}\X_t(u)\right), \quad \eta=1, 2, \dots,n, \label{eq:1}
\end{equation}
where $S_{n,0}^0(u) = S_{n,n}^0(u) = 0$. From~\eqref{eq:1}, we compute the Hilbert-Schmidt norm $\|S_{n, \eta}^{0}(u)\|$. The test statistic is
\[
T_n = \max_{1\leq \eta \leq n}\|S_{n,\eta}^{0}(u)\|^2.
\]
Under the null hypothesis $H_0: \delta=0$, 
\[
T_n \xrightarrow[]{D} \sup_{0\leq x\leq 1}\sum^{\infty}_{\ell=1}\lambda_{\ell}B_{\ell}^2(x), \qquad n\rightarrow \infty,
\]
where $(B_{\ell}:\ell\in \mathcal{N})$ are independent and identically distributed (i.i.d.) standard Brownian bridges defined on $[0,1]$, and $\lambda_{\ell}$ is the $\ell$\textsuperscript{th} eigenvalue of the long-run covariance function specific to functional time series. 

To estimate the long-run covariance function, we utilize a kernel sandwich estimator proposed by \cite{RS17}. This estimator is designed to capture the long-run covariance, which includes autocovariance up to some finite lag. The long-run covariance is defined as
\begin{align*}
C(u, v) &= \sum^{\infty}_{\ell=-\infty}\gamma_{\ell}(u, v) \\
\gamma_{\ell}(u, v) &= \text{cov}[\X(u), \X(v)],
\end{align*}
and is a well-defined element of $\mathcal{L}^2(\mathcal{I})^2$, where $\mathcal{I}$ denotes a compact support interval.

In practice, we need to estimate $C$ from a finite sample $\left\{\X_1(u), \dots, \X_n(u)\right\}$. Given its definition as a bi-infinite sum, a natural estimator of $C$ is
\begin{equation}
\widehat{C}_{h, q}(u, v) = \sum^{\infty}_{\ell=-\infty}W_q\left(\frac{\ell}{h}\right)\widehat{\gamma}_{\ell}(u, v),\label{eq:long-run}
\end{equation}
where $h$ is called the bandwidth parameter,
\[ 
\widehat{\gamma}_{\ell}(u, v) = \left\{ \begin{array}{ll}
         \frac{1}{n}\sum^{n-\ell}_{j=1}[\X_j(u) - \overline{\X}(u)][\X_{j+\ell}(v) - \overline{\X}(v)] & \mbox{if $\ell \geq 0$};\\
        \frac{1}{n}\sum^{n}_{j=1-\ell}[\X_j(u) - \overline{\X}(u)][\X_{j+\ell}(v) - \overline{\X}(v)] & \mbox{if $\ell < 0$}.\end{array} \right. 
\]       
is an estimator of $\gamma_{\ell}(u, v)$ and $W_q(\cdot)$ is a symmetric weight function with bounded support of order $q$. For the kernel estimator, the critical aspect lies in estimating the bandwidth parameter $h$. This can be achieved through a data-driven approach, such as the plug-in algorithm proposed in \cite{RS17}. Computationally, an \Rlogo\ function \verb|long_run_covariance_estimation| is available in the \texttt{ftsa} package \citep{Shang13, HS21}.

We reject $H_0$ if $T_n$ exceeds the corresponding quantile of its distribution. The location of the change point is identified with the largest value of $\|S_{n,\eta}^{0}(u)\|$ for $1\leq \eta \leq n$. In cases of ties, the change point is identified as the first occurrence.

\subsection{Regression-based approach}\label{sec:3.2}

This method considers a functional change-point detection problem from a forecasting perspective. A standard change-point method can be applied to a univariate time series of forecasting errors. We consider a functional principal component regression, where principal component scores are modeled and forecast by a univariate time-series forecasting method \citep{HU07}. 


Building upon \cite{MGG22}, we extend the functional principal component model from a stationary to a nonstationary functional time series. Nonstationarity is indicated by at least one of the principal component scores $(\widehat{\beta}_{t,1},\widehat{\beta}_{t,2},\dots,\widehat{\beta}_{t,K})$ where $K$ is determined using an eigenvalue ratio criterion. We assume that the nonstationary principal component scores follow an $I(1)$ scalar-valued process. Let $r\in \{1,2,\dots,K\}$ denote the number of principal component scores that are $I(1)$ processes. While it is possible for $r=K$, it is more common to observe $r<K$, indicating that the remaining $(K-r)$ processes are stationary \citep[see also][]{CKP16}. In cases where $r<K$, a portion of the underlying process resides in the nonstationary Hilbert space, while the remainder resides in the stationary Hilbert space. This concept relates to co-integration explored in studies such as \cite{BSS17}, \cite{SB19} and \cite{SS22}.

The procedure can be summarized in the following steps.
\begin{enumerate}
\item[1)] Compute the estimated long-run covariance function based on the first-order differenced functional time series, and let $r=\widehat{K}_{\Delta \X}$, where $\widehat{K}_{\Delta \X}$ is the estimated number of functional principal components. We determine $r$ by a modified eigenvalue ratio criterion introduced in \cite{LRS20}. The estimated value of $r$ is determined as the integer that minimizes the ratio of two adjacent empirical eigenvalues, given by
\begin{equation*}
\widehat{r} = \argmin_{1\leq k\leq k_{\max}}\left\{\frac{\widehat{\lambda}_{k+1}}{\widehat{\lambda}_k}\times \mathds{1}\Big(\frac{\widehat{\lambda}_k}{\widehat{\lambda}_1}\geq \theta\Big) + \mathds{1}\Big(\frac{\widehat{\lambda}_k}{\widehat{\lambda}_1}<\theta\Big)\right\},
\end{equation*}
where $k_{\max}$ is a pre-specified positive integer, $\theta$ is a pre-specified small positive number, and $\mathds{1}(\cdot)$ is the binary indicator function. In the absence of prior information, it is reasonable to choose a relatively large $k_{\max}$, for example, $k_{\max}=\#\{k|\widehat{\lambda}_k\geq \sum^n_{k=1}\widehat{\lambda}_k/n, k\geq 1\}$ \citep{AH13}. To prevent over-fitting, we introduce a penalty term for smaller empirical eigenvalues using a threshold constant $\theta=1/\ln[\max(\widehat{\lambda}_1,n)]$, ensuring consistency of $\widehat{r}$.
\item[2)] With the estimated $r$, we compute the estimated functional principal components and their associated principal component scores, $\{\widehat{\zeta}_k\}$ and $\{\widehat{\beta}_{t,k}\}$ for $k=1,2,\dots,r$. The realizations of the stochastic process $\X$ can be expressed as
\begin{equation*}
\X_t(u) = \overline{\X}(u) + \sum_{k=1}^{r}\widehat{\beta}_{t,k}\widehat{\zeta}_k(u),\qquad t=1,2,\dots,n,
\end{equation*}
where $\overline{\X}(u) = \frac{1}{n}\sum^n_{t=1}\X_t(u)$. 
\item[3)] Compute the functional residuals: $\mathcal{Z}_t(u) = \X_t(u) - \overline{\X}(u) - \sum^r_{k=1}\widehat{\beta}_{t,k}\widehat{\zeta}_k(u)$.
\item[4)] Apply \citeauthor{GK07}'s \citeyearpar{GK07} independence test to the residuals $[\mathcal{Z}_1(u),\dots,\mathcal{Z}_n(u)]$. If the $p$-value obtained from the independence test is greater than a level of significance $\alpha=0.05$, then the residuals are deemed independent, and the algorithm terminates. Else, compute the estimated long-run covariance of $[\mathcal{Z}_1(u),\dots,\mathcal{Z}_n(u)]$.
\item[5)] Apply the eigenvalue ratio criterion to determine the optimal number of retained functional principal components $\widehat{K}_{\mathcal{Z}}$.
\item[6)] Obtain all estimated functional principal components and their associated principal component scores: $\{\widehat{\zeta}_1(u),\dots,\widehat{\zeta}_r(u),\widehat{\zeta}_{r+1}(u),\dots, \widehat{\zeta}_{K}(u)\}$ and $\{\widehat{\beta}_{t,1},\dots,\widehat{\beta}_{t,r},\widehat{\beta}_{t,r+1},\dots,\widehat{\beta}_{t,K}\}$ for $t=1,2,\dots,n$ and $K = r+\widehat{K}_{\mathcal{Z}}$.
\item[7)] Obtain forecast scores and forecast curves: using a univariate time-series method, we obtain the principal component score forecast for $h$-step-ahead, $\widehat{\beta}_{n+h|n,k}$. Multiply these forecast scores by the estimated functional principal components to obtain the forecast curves:
\begin{equation*}
\widehat{\X}_{n+h|n}(u) = \overline{\X}(u) + \sum^r_{k=1}\widehat{\beta}_{n+h|n,k}\widehat{\zeta}_k(u)+\sum^{K}_{\nu=r+1}\widehat{\beta}_{n+h|n,\nu}\widehat{\zeta}_{\nu}(u).
\end{equation*}
\end{enumerate}

In the forecasting procedure, we begin by generating one-step-ahead point forecasts using the first three functional observations. Subsequently, we expand the sample to include the first four functional observations and re-estimate the parameters within a functional time-series model. We then generate forecasts one step ahead based on the estimated models. This iterative process continues, incrementally increasing the sample size by one observation, until reaching the end of the sampling period. This approach results in 98 one-step-ahead forecasts. To assess point forecast accuracy, we compare these forecasts with the holdout samples (i.e., curves from the 4\textsuperscript{th} to the 101\textsuperscript{st} observations).

To evaluate the accuracy of point forecasts, we employ the integrated squared forecast error (ISFE) \citep[see also][]{HU07}. This metric quantifies the proximity of forecasts to the actual values of the forecasted variable, and is defined as
\begin{equation*}
\kappa_{\gamma+1} = \int_{\mathcal{I}} [\X_{\gamma+1}(u) - \widehat{\X}_{\gamma+1|\gamma}(u)]^2du, \qquad \gamma = 3, \dots, (n-1),
\end{equation*}
where $\X_{\gamma+1}$ represents the $(\gamma+1)$\textsuperscript{th} holdout sample in the forecasting scheme, and $\widehat{\X}_{\gamma+1|\gamma}(u)$ denotes the one-step-ahead point forecasts for the holdout sample.

Since our objective is to identify the optimal fitting period with one breakpoint \citep[see also][]{BMS02}, we estimate the breakpoint using the methodology outlined by \cite{BP03} (see \cite{ZKK+03} and \cite{ZK05} for detailed implementation descriptions). Suppose we have a time series of integrated squared forecast errors $\kappa_{\gamma+1}$ for $\gamma=3,\dots,(n-1)$. We estimate a random walk with a piecewise constant drift for the time-dependent variables:
\begin{equation*} 
\Delta \kappa_{\gamma+1} = \left\{ \begin{array}{ll}
\varsigma_{1} + \xi_{\gamma+1} & \mbox{\qquad$\gamma+1 \leq \eta^{*}$}\\
\varsigma_{2}+\xi_{\gamma+1} & \mbox{\qquad$\gamma+1 > \eta^{*}$}\end{array} \right. 
\end{equation*}
where $\varsigma_{1}$ and $\varsigma_{2}$ denote the respective mean terms before and after a breakpoint, and $\xi_{\gamma+1}$ represents the error term. We estimate this model using ordinary least squares by minimizing the sum of squared residuals (SSR):
\begin{equation*}
\text{SSR}\left(\gamma^*\right) = \sum_{\gamma=3}^{\gamma^*-1}\left(\Delta \kappa_{\gamma+1} - \varsigma_{1}\right)^2 + \sum_{\gamma=\gamma^*}^{n-1}\left(\Delta \kappa_{\gamma+1} - \varsigma_2\right)^2.
\end{equation*}

This model specification identifies one breakpoint, dividing the univariate time series of forecast errors into two regimes with distinct shifts. The optimal stopping time is selected to minimize the SSR.

\section{Simulation studies}\label{sec:6}

We consider three data generating processes previously studied in \cite{SCS22}. The first one considers a stationary functional time series in Section~\ref{sec:6.1}, where there is an abrupt change occurs at a pre-fixed location. In Sections~\ref{sec:6.2} and~\ref{sec:6.3}, we consider a nonstationary functional time series, where abrupt and gradual changes take place at a randomly assigned location, respectively.

\subsection{An abrupt change in the mean of a stationary functional time series}\label{sec:6.1}

Via a series of Monte-Carlo simulation studies, we evaluate the performance of the two detection methods. The data generating process is a pointwise FAR(1), given by
\begin{align*}
\X_1(u) &= 10\times u\times (1-u) + \omega \times B_1(u) \\
\X_t(u)  &= (\rho+c)\X_{t-1}(u) + \omega \times B_t(u), \quad t=2,\dots,n \\
\Y_t(u) &= \frac{|\X_{t-1}(u) - \X_t(u)|}{|\X_{t-1}(u)+0.1|},
\end{align*}
where $\{B_t(u), t=1,\dots,n, u\in [0,1]\}$ denotes i.i.d. standard Brownian motions. In practice, we discretize continuum $u$ on 101 equally spaced grid points. We consider three values of $\omega=0.1, 0.5, 0.9$ to reflect three levels of noise to signal. The coefficients satisfy $|\rho|<1$ and $|\rho+c|<1$ in order to ensure stationarity. Let $\rho=0.2$ and $c=0$ for curves from two to $\tau=\lceil n/2 \rceil$; while $c=0.7$ for curves from $\lceil n/2\rceil+1$ to $n$. 

\begin{table}[!htb]
\centering
\tabcolsep 0.14in
\caption{\small{For three sample sizes, we determine the mean, median, standard deviation, and mean squared error of the estimated change points obtained from the two detection methods.}}\label{tab:6.1}
\begin{small}
\begin{tabular}{@{}lllcccccccc@{}}
\toprule
	&		&		& \multicolumn{8}{c}{$\widehat{\tau}$} \\
	&		&		&  \multicolumn{4}{c}{Fully functional} & \multicolumn{4}{c}{Regression based} \\
	\cmidrule{4-11}
$n$ & $\tau$ & $\omega$ & mean & median & sd & MSE & mean & median & sd & MSE  \\
\midrule
101 	& 51 		& 0.1 & 50.27   & 51   & 7.70 & 59.28 & 51.86   & 52   & 17.73 & 317.61   \\
	&		& 0.5 & 52.39   & 52   & 5.87 & 40.09 & 50.80   & 52   & 18.54 & 343.91 	\\
	&		& 0.9 & 52.45   & 52   & 5.76 & 39.19 & 50.48   & 52   & 18.45 & 340.12	\\
201 	& 101 	& 0.1 & 100.60 & 101 & 8.76 & 77.02 & 100.53 & 102 & 30.57 & 934.08	\\
	&	  	& 0.5 & 102.53 & 101 & 6.84 & 53.18 & 98.08   & 101 & 32.03 & 1028.34 	\\
	&	  	& 0.9 & 102.72 & 101 & 6.72 & 52.57 & 98.19   & 101 & 30.73 & 946.75	\\
401	& 201 	& 0.1 & 201.25 & 201 & 9.82 & 97.88 & 197.98 & 201 & 45.36 & 2059.12 	\\
	&	  	& 0.5 & 203.58 & 202 & 8.32 & 81.89 & 195.89 & 201 & 47.55 & 2275.74 	\\
	&	  	& 0.9 & 203.59 & 202 & 8.12 & 78.76 & 194      & 200 & 46.38 & 2185.38 	\\
\bottomrule
\end{tabular}
\end{small}
\end{table}

\subsection{An abrupt change in the mean of a nonstationary functional time series}\label{sec:6.2}

We consider another data generating process for simulating functional time series. We begin with simulating a time series of error functions $[\epsilon_1(u), \epsilon_2(u), \dots, \epsilon_{n}(u)]$ given below:
\begin{equation*}
\epsilon_{t}(u) = \sum^{K}_{k=1}\beta_{t,k}\phi_{k}(u) + \varepsilon_{t}(u),
\end{equation*}
where $\left[\phi_{1}(u),\phi_{2}(u),\dots,\phi_{K}(u)\right]$ are randomly sampled with replacement from $K=21$ Fourier basis functions, and $\varepsilon_{t}(u)$ denotes an innovation term that can be independent over $t$. We consider 101 equally spaced grids between 0 and 1.

Let $\bm{\beta}_t = (\beta_{t,1},\beta_{t,2},\dots,\beta_{t,K})^{\top}$ be a $K$-dimensional vector. We generate $\bm{\beta}_{t}$ from a vector autoregressive of order 1 (VAR(1)) model,
\begin{equation*}
\bm{\beta}_t = \bm{A}\bm{\beta}_{t-1}+\bm{\psi}_t, \qquad t=2,\dots,n,
\end{equation*}
where $\bm{A}=(a_{ij})_{K\times K}$ is the VAR(1) coefficient matrix, and $\bm{\psi}_t$ denotes the error terms of the VAR(1) model at time $t$. Following \cite{LRS20}, we consider two structures for coefficient matrix $\bm{A}$.
\begin{enumerate}
\item[(1)] $\bm{A}$ is a diagonal matrix with diagonal elements drawn from a $U(-0.5, 0.5)$ and $\bm{\psi}_{t}$ is generated by a $K$-dimensional normal distribution with mean zero and power-decay covariance structure $\text{cor}(\psi_t^i, \psi_t^j) = \rho^{|i-j|}$, where $\rho$ denotes a correlation parameter, such as $\rho=0.5$.
\item[(2)] Alternatively, $\bm{A}$ is a banded autoregressive matrix with $a_{i,j}$ independently drawn from a $U(-0.3, 0.3)$ when $|i-j|\leq 3$ and $a_{ij}=0$ when $|i-j|>3$, and $\bm{\psi}_t$ is independently generated by a $K$-dimensional normal distribution with mean zero and identity covariance matrix.
\end{enumerate}

To specify a change-point location for the population, we draw a value from a $U(0.25\times n, 0.75\times n)$. The lower and upper bounds of the uniform distribution are chosen so that the location of a change point does not lie on the boundary of a sample. 

Following \cite{ARS18}, a class of break functions was given by
\begin{align*}
\delta_{k}^{*}(u) &= \frac{1}{\sqrt{k}}\sum^{k}_{w=1}\phi_{w}(u),\quad k=1,2,\dots,K, \\
\delta_{k}(u) &= \delta_{k}^{*}(u)\times \sqrt{c},
\end{align*}
where the normalization is required to ensure $\delta_k^*(u)$ has unit norm. Here, we consider $\delta_1(u)$ in which the break occurs in the leading eigendirection. The value of $c$ controls the magnitude of the break, and it links to the signal-to-noise (SNR) ratio
\begin{equation*}
\text{SNR} = c\times \frac{\mathrm{p}\times (1-\mathrm{p})}{\text{tr}(\widehat{C}_{\epsilon})},
\end{equation*}
where $\text{tr}(\widehat{C}_{\epsilon})$ denotes the trace of the estimated long-run covariance of the error term. For a given SNR value, we can, in turn, compute the corresponding value of $c$.

With a chosen eigendirection, such as $k=1$, we simulate $n$ samples of a non-stationary functional time series as follows:
\begin{align*}
\X_t(u) &= \delta_1(u)\times \mathds{1}\{t>\tau\}+\epsilon_t(u), \\
\Y_t(u) &= \overline{\X}(u) + \X_t(u),
\end{align*}
where $\overline{\X}(u) = \frac{1}{n}\sum^n_{t=1}\X_t(u)$.

Table~\ref{tab:2} presents some summary statistics of our estimated change points obtained from the two detection methods for three sample sizes for four different SNR ratios. As SNR increases from 0.01 to 0.9, the MSE values decrease, implying that both detection methods make locating the actual change points easier. The fully functional detection method achieves zero MSE values.
\begin{table}[!htb]
\centering
\begin{small}
\tabcolsep 0.08in
\caption{\small{For three sample sizes, we determine the mean and median of the actual change points, and mean, median, sd, and MSE of the estimated change points under four different signal-to-noise ratios.}\label{tab:2}}
\begin{tabular}{@{}llllrrrrrrrr@{}}
\toprule
& & \multicolumn{2}{c}{True change point $\tau$} & \multicolumn{8}{c}{Estimated change point $\widehat{\tau}$} \\
& & 	& &			 \multicolumn{4}{c}{Fully functional} & \multicolumn{4}{c}{Regression based} \\
$n$ & SNR & mean & median & mean & median & sd & MSE  & mean & median & sd & MSE \\
\midrule
\multicolumn{8}{l}{\hspace{-.1in}{\underline{$\bm{A}$ = Band}}} \\
100 & 0.01 & 50.07 & 50 & 50.07 & 50 & 14.20 & 0 & 51.21 & 52 & 18.87 & 348.48\\
	& 0.1 &  &  	& 50.07 & 50 & 14.20 & 0 & 51.58 & 52 & 12.95 & 5.22 \\
	& 0.5 &  &		 & 50.07 & 50 & 14.20 & 0 & 51.57 & 52 & 12.93 & 4.52 \\
	& 0.9 &  &		 & 50.07 & 50 & 14.20 & 0 & 51.57 & 52 & 12.93 & 4.50 \\
\midrule
200 	& 0.01 & 99.99 & 99 & 99.99 & 99 & 29.41 & 0 & 99.74 & 99 & 40.64 & 1897.16 \\
	& 0.1   &  &  & 99.99 & 99 & 29.41 & 0 & 101.42 & 100 & 27.90 & 13.22 \\
	& 0.5   &  &  & 99.99 & 99 & 29.41 & 0 & 101.52 & 101 & 28.13 & 4.61 \\
	& 0.9   &  &  & 99.99 & 99 & 29.41 & 0 & 101.52 & 101 & 28.13 & 4.59 \\ 
\midrule
400  & 0.01 & 200.83 & 203.50 & 200.83 & 203.50 & 58.36 & 0 & 200.01 & 206 & 88.73 & 9403.59 \\
	& 0.1	   & & & 200.83 & 203.50 & 58.36 & 0 & 202.69 & 206.50 & 55.93 & 140.63 \\
	& 0.5   &  & & 200.83 & 203.50 & 58.36 & 0 & 202.28 & 205.50 & 57.04 & 4.64 \\
	& 0.9   &  & & 200.83 & 203.50 & 58.36 & 0 & 202.29 & 205.00 & 57.06 & 4.42 \\
\midrule
\multicolumn{8}{l}{\hspace{-.1in}{\underline{$\bm{A}$ = Diag}}} \\
100 & 0.01 & 49.46 & 49 & 49.46 & 49 & 14.74 & 0 & 51.54 & 52 & 19.24 & 353.91\\
	& 0.1 &   &  & 49.46 & 49 & 14.74 & 0 & 51.04 & 51 & 13.44 & 5.94 \\
	& 0.5 & &  & 49.46 & 49 & 14.74 & 0 & 51.02 & 51 & 13.45 & 4.70 \\
	& 0.9 &  &  & 49.46 & 49 & 14.74 & 0 & 51.02 & 51 & 13.45 & 4.70 \\
\midrule
200	& 0.01 & 99.6 & 100 & 99.6 & 100 & 28.54 & 0 & 103.05 & 105 & 41.35 & 2007.33 \\
	& 0.1 &  & & 99.6 & 100 & 28.54 & 0 & 101.18 & 102 & 26.96 & 13.21 \\
	& 0.5 & &  & 99.6 & 100 & 28.54 & 0 & 101.11 & 101 & 27.25 & 4.56 \\
	& 0.9 & & & 99.6 & 100 & 28.54 & 0 & 101.11 & 101 & 27.25 & 4.54 \\
\midrule
400 & 0.01 & 200.39 & 199.5 & 200.39 & 199.5 & 59.03 & 0 & 199.57 & 205 & 87.46 & 9899.3 \\
	& 0.1 & &  & 200.39 & 199.5 & 59.03 & 0 & 201.36 & 202 & 57.06 & 125.59 \\
	& 0.5 &  &  & 200.39 & 199.5 & 59.03 & 0 & 201.92 & 201 & 57.72 & 4.98 \\
	& 0.9 &  & & 200.39 & 199.5 & 59.03 & 0 & 201.91 & 201.5 & 57.73 & 4.58 \\	
\bottomrule
\end{tabular}
\end{small}
\end{table}

\subsection{A gradual change in the mean of a nonstationary functional time series}\label{sec:6.3}

While Sections~\ref{sec:6.1} and~\ref{sec:6.2} study stationary and nonstationary functional time series with an abrupt change, in Section~\ref{sec:6.2}, we alter the data generating process from a sudden change to a gradual change in mean. With the chosen eigendirection $k=1$, we simulate $n$ samples of a non-stationary functional time series as follows:
\begin{align*}
\X_t(u) &= \sqrt{t}\times \frac{n^{\alpha}}{\sqrt{n}}\times \delta_1(u)\times \mathds{1}\{t>\tau\}+\epsilon_t(u), \\
\Y_t(u) &= \overline{\X}(u) + \X_t(u),
\end{align*}
where $t$ is a time-varying index representing a gradual change. The value of $\alpha\in (0, \frac{1}{2})$ is a constant, along with $c$ controlling the magnitude of the change point. As $\alpha$ or $c$ value increases, the change point size increases; thus it is easier to detect them. In our simulation study, we set $\alpha=0.5$.

In Table~\ref{tab:3}, we present some summary statistics of our estimated change points obtained from the two detection methods for three sample sizes for four different SNR ratios. As SNR increases from 0.01 to 0.9, the MSE values decrease, implying that both detection methods make locating the actual change points easier. Note that when SNR=0.5 or 0.9, the regression-based detection method outperforms the fully functional detection methods for $n=200$ or 400.

\begin{center}
\begin{small}
\begin{longtable}{@{}llllrrrrrrrr@{}}
\caption{\small{For three sample sizes, we determine the mean and median of the actual change points, and mean, median, sd, and MSE of the estimated change points under four different signal-to-noise ratios.}\label{tab:3}}\\
\toprule
& & \multicolumn{2}{c}{True change point $\tau$} & \multicolumn{8}{c}{Estimated change point $\widehat{\tau}$} \\
& & 	& &			 \multicolumn{4}{c}{Fully functional} & \multicolumn{4}{c}{Regression based} \\
$n$ & SNR & mean & median & mean & median & sd & MSE  & mean & median & sd & MSE \\
\midrule
\endfirsthead
\toprule
& & \multicolumn{2}{c}{True change point $\tau$} & \multicolumn{8}{c}{Estimated change point $\widehat{\tau}$} \\
& & 	& &			 \multicolumn{4}{c}{Fully functional} & \multicolumn{4}{c}{Regression based} \\
$n$ & SNR & mean & median & mean & median & sd & MSE  & mean & median & sd & MSE \\
\midrule
\endhead
\midrule
\multicolumn{12}{r}{{Continued on next page}} \\
\endfoot
\endlastfoot
\multicolumn{8}{l}{\hspace{-.1in}{\underline{$\bm{A}$ = Band}}} \\
100 & 0.01 & 50.07 & 50 & 50.79 & 50 & 13.19 & 5.42 & 52.96 & 55 & 19.68 & 416.13 \\
	& 0.1  & & & 50.63 & 50 & 13.36 & 3.67 & 51.65 & 52 & 12.94 & 8.17 \\
	& 0.5 &  &  & 50.63 & 50 & 13.36 & 3.66 & 51.54 & 51 & 12.94 & 4.42 \\
	& 0.9 &  &  & 50.62 & 50 & 13.37 & 3.61 & 51.55 & 51 & 12.93 & 4.45 \\
\midrule
200	 & 0.01 & 99.99 & 99 & 101.42 & 99 & 27.47 & 21.54 & 103.46 & 106 & 41.38 & 2041.04 \\
	& 0.1 &  &  & 101.22 & 99 & 27.68 & 16.95 & 101.89 & 101 & 27.62 & 33.27 \\
	& 0.5 &  &  & 101.15 & 99 & 27.76 & 15.24 & 101.48 & 101 & 28.13 & 4.52 \\
	& 0.9 &  &  & 101.16 & 99 & 27.75 & 15.25 & 101.50 & 101 & 28.13 & 4.53 \\
\midrule
400 & 0.01 & 200.83 & 203.5 & 203.53 & 203.5 & 54.52 & 74.31 & 204.53 & 217 & 88.91 & 9622.83 \\
	& 0.1 &   &  & 203.30 & 203.5 & 54.78 & 63.44 & 204.57 & 209 & 55.23 & 302.56 \\
	& 0.5 &   &  & 203.25 & 203.5 & 54.84 & 61.75 & 202.34 & 205.5 & 57 & 5.04 \\
	& 0.9 &   &  & 203.23 & 203.5 & 54.86 & 61.11 & 202.29 & 204.5 & 57.06 & 4.41 \\
\midrule
\multicolumn{8}{l}{\hspace{-.1in}{\underline{$\bm{A}$ = Diag}}} \\
100 & 0.01 & 49.46 & 49 & 50.22 & 49 & 13.74 & 5.75 & 54.05 & 56 & 19.67 & 444.04 \\
	& 0.1 & 	&	& 50.08 & 49 & 13.87 & 4.07 & 51.21 & 51 & 13.35 & 9.51 \\
	& 0.5 &	&	& 50.08 & 49 & 13.87 & 4.04 & 51.02 & 51 & 13.46 & 4.68 \\
	& 0.9 &	&	& 50.08 & 49 & 13.87 & 3.98 & 51.02 & 51 & 13.45 & 4.68 \\
\midrule
200 & 0.01 & 99.6 & 100 & 100.82 & 100 & 26.84 & 15.45 & 103.69 & 108 & 42.65 & 2050.37 \\
	& 0.1	 &	&	& 100.61 & 100 & 27.07 & 11.94 & 101.65 & 102 & 26.70 & 28.86 \\
	& 0.5 &	&	&100.60	& 100	& 27.09 & 12.26 & 101.11 & 101 & 27.24 & 4.62 \\
	& 0.9 & 	&	&100.59 & 100	& 27.10 & 12.03 & 101.10 & 101 & 27.25 & 4.53 \\
\midrule
400 & 0.01 & 200.39 & 199.5 & 202.83 & 199.5 & 55.68 & 64.57 & 203.48 & 214.5 & 88.25 & 9900.08 \\
	& 0.1 & 	&	& 202.51 & 199.5 & 56.04 & 52.28 & 203.07 & 205 & 56.97 & 285.37 \\
	& 0.5 &	&	& 202.43 & 199.5 & 56.13 & 48.83 & 201.97 & 201 & 57.68 & 5.40 \\
	& 0.9 & 	& 	& 202.44 & 199.5 & 56.12 & 48.73 & 201.90 & 201 & 57.73 & 4.66 \\
\bottomrule
\end{longtable}
\end{small}
\end{center}

\section{Application to age-specific mortality and fertility rates}\label{sec:4}

\subsection{Change-point detections}

We applied two change-point detection methods to identify the location of the change point. Using the fully functional method, we found change points in~1972 and~1977 for Australian female and male mortality, respectively. The regression-based approach indicated change points in~1976 and~1982. For comparison, implementing the approach by \citeauthor{BMS02} (\citeyear{BMS02}) identified change points in~1981 and~1973.

In the Australian age-specific fertility rates, the fully functional detection method identified a change point in~1975. Using the regression-based approach, the detected change point was in~1991. For comparison, implementing the approach by \citeauthor{BMS02} (\citeyear{BMS02}) identified a change point in~1974.

\subsection{The relationship between the fitting period and forecast accuracy}

We investigate how the fitting period influences forecast accuracy by dividing the dataset into training and testing samples. The testing sample comprises the last ten years of data, while the initial training sample includes the remaining data. We employ two change-point detection methods to locate the change point and define the start of the fitting period. Focusing on the LC method, we model and forecast age-specific $\log_{10}$ mortality and fertility rates. The forecasting method remains consistent across experiments, with the sole variation being the length of the fitting period.

For the Australian age-specific fertility rates, we begin with an initial training sample spanning from 1921 to 2011. Using this data, we generate a one-step-ahead forecast for the year 2012. Subsequently, we adopt an expanding window approach, incrementally increasing the training sample by one year and generating one-step-ahead forecasts for each subsequent year up to 2021. This iterative process yields ten one-step-ahead forecasts in total. To evaluate the accuracy of these forecasts, we compare them with the holdout sample from 2012 to 2021.

We assess the point forecast accuracy using the mean absolute percentage error (MAPE), defined as
\begin{equation*}
\text{MAPE} = \frac{1}{J\times 10}\sum_{j=1}^J\sum_{\varphi=1}^{10}\frac{\left|\X_{m+\varphi}(u_j) - \widehat{\X}_{m+\varphi|m+\varphi-1}(u_j)\right|}{\X_{m+\varphi}(u_j)},
\end{equation*}
where $\X_{m+\varphi}(u_j)$ denotes the $(m+\varphi)$\textsuperscript{th} holdout curve at a discrete point $u_j$, and $\widehat{\X}_{m+\varphi|m+\varphi-1}(u_j)$ denotes its corresponding forecast.

Using the initial training sample, we apply two change-point detection methods to identify pivotal years. The regression-based approach identifies 1992 as significant, while the fully functional method pinpoints 1974. We then truncate our initial training sample from these identified change points to their respective periods. Subsequently, we employ the LC method to forecast age-specific fertility rates in the testing period.

\begin{figure}[!htb]
\centering
\includegraphics[width=8.4cm]{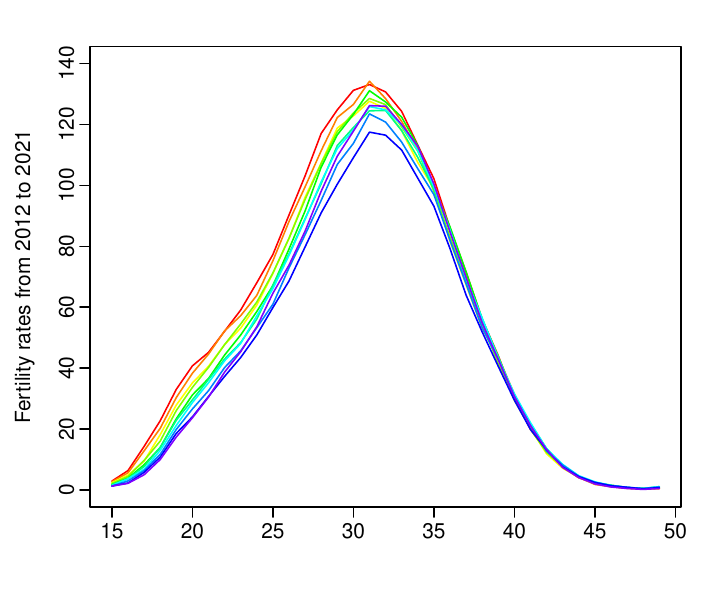}
\quad
\includegraphics[width=8.4cm]{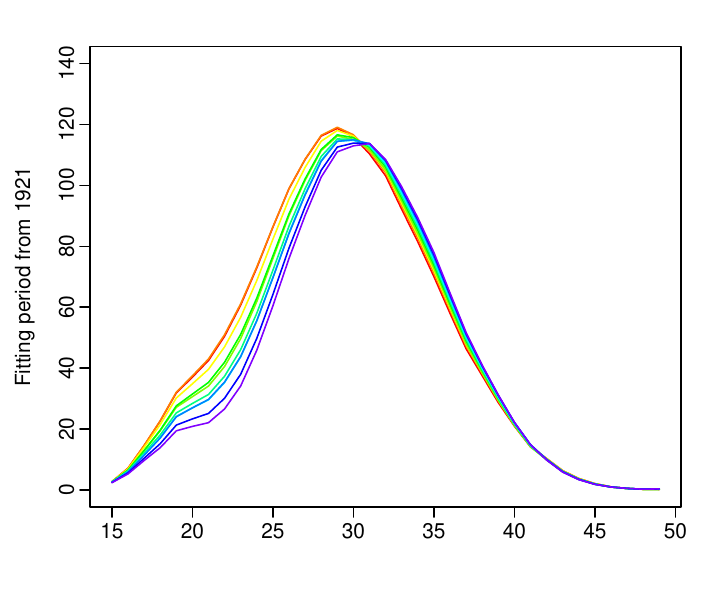}
\\
\includegraphics[width=8.4cm]{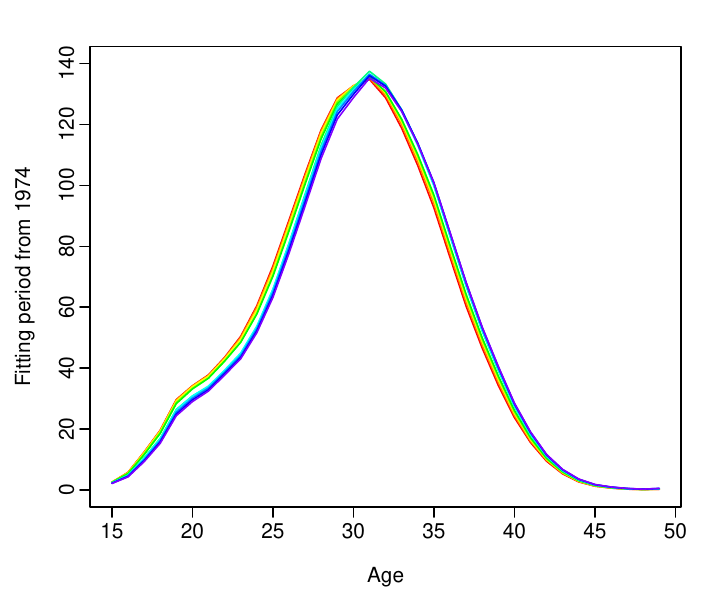}
\quad
\includegraphics[width=8.4cm]{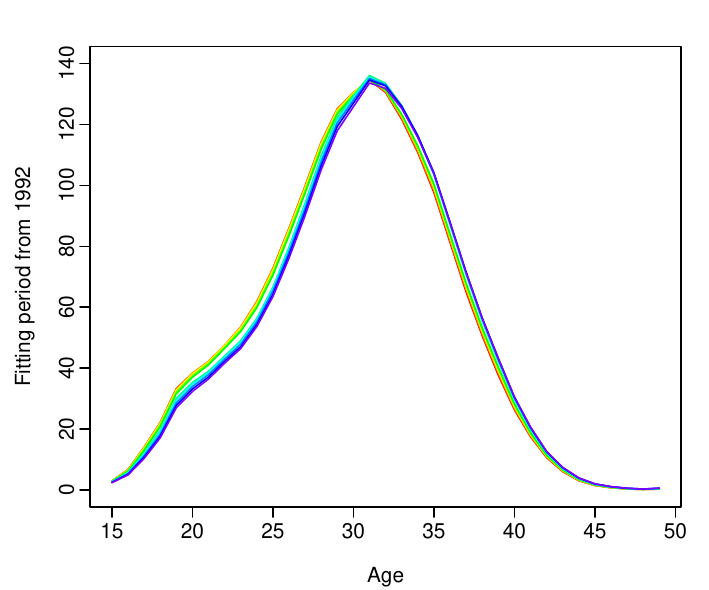}
\caption{\small{Holdout sample and forecasts of the Australian age-specific fertility rates between 2012 and 2021.}}\label{fig:3}
\end{figure}
Figure~\ref{fig:3} illustrates comparisons between the holdout sample and forecasts generated from different fitting periods. By juxtaposing the holdout fertility data against these forecasts, we compute the mean absolute percentage error (MAPE) as shown in Table~\ref{tab:fert_MAPE}.

\begin{table}[!htb]
\tabcolsep 0.64in
\centering
\caption{\small{A comparison of MAPE between the holdout data and the forecasts. Using the LC method, we consider several change-point detection methods for determining the optimal training sample.}}\label{tab:fert_MAPE}
\begin{tabular}{@{}llll@{}}
\toprule
		& \multicolumn{2}{c}{Mortality} 	& Fertility \\
Method 	& Female 	& Male 			& \\
\midrule
Full sample  						 & 3.36 & 4.16 & 22.69 \\
Fully functional 						 &  1.94 & 2.27 & 17.42 \\
Regression-based 					 & 1.81 & 1.96 & 15.37 \\ 
\citeauthor{BMS02}'s \citeyearpar{BMS02} & 1.88 & 2.22 & 17.47 \\
\bottomrule
\end{tabular}
\end{table}

We divide the Australian age-specific female and male mortality rates into an initial training sample spanning from 1921 to 2010, with a subsequent testing sample covering 2011 to 2020. Employing two change-point detection methods, we identify change points in the female series in 1960 (fully functional approach) and 1974 (regression-based approach), while for the male series, change points are identified in 1974 (fully functional approach) and 1982 (regression-based approach). Using the LC method, we generate ten one-step-ahead forecasts of $\log_{10}$ mortality rates and assess their accuracy using the MAPE.

\section{Conclusion}\label{sec:5}

This paper has focused on two change-point detection methods to find breaks in mean functions of functional time series, which exhibit both cross-sectional correlations across ages and temporal dependence. The first method employs a scaled functional cumulative sum statistic to pinpoint the change point, identified as the first location with the highest cumulative sum statistic value. The second method reformulates change-point detection as a forecasting challenge, determining the change point where it yields the largest one-step-ahead forecast errors. Via a series of simulation studies, we evaluate and compare the finite-sample performance of the two detection methods and recommend the fully functional method. Using Australian age-specific mortality and fertility rates data, we successfully identify change points and determine the optimal starting year for the fitting period. By comparing forecasts based on different calibration periods, our findings indicate that reducing the fitting period enhances forecast accuracy. To facilitate reproducibility, the \Rlogo \ code is available at \url{https://github.com/hanshang/change_point}.

There are several ways in which the methods presented can be extended, and we briefly list four. First, we study several change-point detection methods to locate the change point and find the optimal training sample period. An alternative approach is to adopt the model averaging idea to select the optimal training period implicitly \citep[see, e.g.,][Chapter 3]{Kessy22}. Second, we demonstrate how the optimal training sample period can affect point forecast accuracy. The presence of a change point can also affect the interval forecast accuracy. Third, we aim to identify a change point for a single functional time series. It is also valuable to explore change-point detection methods for subnational demographic curves, where each series may exhibit a change point. By combining all change points, one can also identify common change points \citep[see, e.g.,][]{LLS24}. Fourth, we choose to locate change points in the age-specific mortality and fertility rates. Alternatively, one can also study the changes in the proportion of causes of death over time \citep[see, e.g.,][]{NBC+24}.

\section*{Conflict of interest}

The author declares no competing interests.

\section*{Acknowledgements}

The author thanks two reviewers, whose valuable comments led to an improved version of this article.

\bibliographystyle{agsm}
\bibliography{strucchange.bib}

\end{document}